# Design of nanoparticles for generation and stabilization of $CO_2$-in-brine foams with or without added surfactants

Andrew J. Worthen[†], Shehab Alzobaidi[†], Vu Tran[†], Muhammad Iqbal[†], Jefferson S. Liu[†], Kevin A. Cornell[†], Ijung Kim[‡], David A. DiCarlo[‡], Steven L. Bryant[†‡], Chun Huh[‡], Thomas M. Truskett[†], and Keith P. Johnston[†].

[†]McKetta Department of Chemical Engineering, University of Texas at Austin, United States
[‡]Department of Petroleum & Geosystems Engineering, University of Texas at Austin, United States
[†‡]Department of Chemical & Petroleum Engineering, University of Calgary, Canada

ABSTRACT

Whereas many studies have examined stabilization of emulsions and foams in low salinity aqueous phases with nanoparticles (NPs) with and without added surfactants, interest has grown recently in much higher salinities relevant to subsurface oil and gas applications. It is shown for the first time that NPs grafted with well-defined low molecular weight ligands colloidally stable in concentrated brine (in particular, API brine, 8% NaCl + 2% $CaCl_2$) and are interfacially active at the brine-air interface. These properties were achieved for three types of ligands: a nonionic diol called GLYMO and two short poly(ethylene glycol) (PEG) oligomers with 6-12 EO repeat units. Carbon dioxide-in-water (C/W) foams could be formed only with modified NPs with higher surface pressures at the A/W interface. Furthermore, these ligands were sufficiently $CO_2$-philic that the hydrophilic/$CO_2$-philic balance of silica NPs was low enough for stabilization of $CO_2$-in-water (C/W) foam with API brine. Additionally, NPs with these three ligands formed stable dispersions with various free molecular surfactants in DI water and even API brine (8% NaCl + 2% $CaCl_2$) at room temperature. A wide variety of mixtures of NPs plus anionic, nonionic, or cationic mixtures that formed stable dispersions were also found to stabilize C/W foams in porous media at high salinity. These results provide a basis for future studies of the mechanism of foam stabilization with NPs and NP/surfactant mixtures at high salinity.

**INTRODUCTION**

Nanoparticles (NPs) have been investigated as an alternative to surfactants as amphiphiles to stabilize emulsions and foams for many applications. For a 10 nm nanoparticle with a contact angle of 90° at an oil-water interface, the adsorption energy is of order $10^3$ kT, relative to only about $10^0$ kT for molecular surfactants, which are much smaller.[1] NPs are desirable because of their chemical stability especially at elevated temperatures, low retention on mineral surfaces, and ability to produce highly stable foams.[2-3] Adsorbed NPs at fluid-fluid interfaces impart stability to foams and emulsions by inhibiting destabilization mechanisms such as drainage of the liquid in the lamella, coalescence of neighboring bubbles (lamella rupture), and Ostwald ripening.[4] Covalently grafted permanent ligands on NP surfaces may be designed to provide colloidal stability in harsh reservoir environments at elevated temperatures and salinities. Of particular interest are new, low molecular weight (MW) coatings containing a sulfobetaine group, short PEG oligomers with 6-12 EO repeat units, and a nonionic diol called "GLYMO" which were demonstrated to stabilize ca. 10~100 nm diameter silica NPs in high salinity API brine (8% NaCl + 2% $CaCl_2$) at up to 80°C for over 30 days,[5] although their interfacial activity was not studied.

Recently, NPs have gained attention in emerging large scale applications of enhanced oil recovery (EOR)[2, 6-7] and hydraulic fracturing,[8] particularly as $CO_2$-in-water foam (or emulsion) stabilizers.[9] For example, silica NPs with the appropriate amount of grafted dimethyl groups on their surfaces have been demonstrated to be sufficiently $CO_2$-philic to provide a sufficiently-low hydrophilic/$CO_2$-philic balance (HCB)[10] to adsorb at a $CO_2$-DI water interface and stabilize C/W emulsions[11] (~50% $CO_2$:water by volume) and foams[2] (>74% $CO_2$ by volume). These foams underwent minimal coalesce after 23 h at 50°C.[2] NPs modified with various degrees of dimethyl groups have also been used to stabilize oil-in-water emulsions[12] and air-in-water foams.[13] A surface pressure of ca. 20 mN/m was observed at an air-water interface[14] indicating their interfacial activity. However, hydrophobic ligands such as dimethyl groups do not provide colloidal stability in high salinity aqueous phases often encountered in subsurface applications.[5] Recently, NPs with unknown coatings have been reported for stabilizing foams at high salinities. For example, "short-chain" poly(ethylene glycol) (PEG)-coated silica NPs stabilized viscous and opaque white C/W foams with up to 8% NaCl in the aqueous phase[2] and silica NPs with an unknown coating stabilized C/W foams at up to 2% KCl.[6, 8, 15] However, the design of NPs with known compositions and amounts of ligands on the surface which can stabilize viscous C/W foams in high salinity aqueous phases remains an elusive goal.

The literature provides various examples of NPs with interfacially-active polymeric coatings for air-water and oil-water systems, which may serve as a basis for designing ligands for C/W foam formation. For example, Stefaniu and coworkers demonstrated monodisperse iron oxide nanoparticles with catechol-terminated brush polymers of either 2-(2-methoxyethoxy) ethyl methacrylate ($MEO_2MA$)[16-17] or both $MEO_2MA$ and oligo(ethylene glycol) methacrylate (OEGMA)[18-19] adsorb irreversibly at the air-water interface and lower the surface tension from 72 to 43~47 mN/m with as little as ca. 0.0003 % w/v amphiphile. Saleh and coworkers[20] used grafted poly(styrenesulfonate) brushes on silica NPs to lower the interfacial tension between trichloroethylene and water from 30 to 15 mN/m with 0.1 w/v % hybrid particles, and O/W emulsions were stable for 6 months with only 0.04 w/v % particles.[20] In addition, Alvarez et al.[21] found silica NP cores with poly((2-(dimethylamino) ethyl methacrylate) brushes "grafted from" the surface reduced $\gamma$ from 38 to 9 mN/m at a concentration of 0.05 w/v %, building on an earlier study by Saigal et al.[22] All of these examples contain surface active polymers with distinct hydrophilic and hydrophobic regions. Here, high MWs are beneficial as it allows the polymer to spread and orient its various regions at the fluid-fluid interface to more efficiently lower interfacial tension.[19, 23] However, polymeric coatings are often expensive due to their high mass fraction in the resulting polymer/particle hybrid material, generally 20+%. Therefore, design of low MW ligands which provide interfacial activity and colloidal stability in harsh subsurface conditions would be highly desirable.

Another concept for raising the interfacial activity of NPs is to mix the NPs with surfactants to modify their surfaces.[24-27] However, if the NP-surfactant attraction is too strong, it may lead to flocculation of the NPs. A modest degree of flocculation of NPs can improve emulsion[25-26] or foam[27] stabilization, but it can be highly undesirable for subsurface applications because the flocs may not pass through porous media. Most design strategies rely on the modification of the surfactant structure to tune the strength of the NP-surfactant interactions.[24-27] An alternative approach would be to modify a NP surface through covalent grafting of ligands to tune the interactions between the modified NPs and molecular surfactant. This approach could potentially be used to control the colloidal stability and to provide synergy in emulsion or foam formation, but to our knowledge is has not been utilized at high salinities to date. Our recent development of ligands that provide colloidal stability to NPs at extremely high salinities relevant to subsurface applications (e.g. API brine described by Worthen et al.[5]) may serve as a basis for the design of new stable NP-surfactant combinations for foam formation. However, very little is known about the phase behavior of mixtures of NPs and surfactants at high salinities and how they influence interfacial properties and the formation of C/W foams.

The objective of this study is to design silica NPs with grafted low MW ligands that stabilize C/W foam in porous media at salinities up to the high levels in API brine and at temperatures up to 50°C. The primary focus is on systems stabilized by NPs alone, whereas the secondary interest is on mixtures of modified NPs and ungrafted molecular surfactants. The design of the particle surfaces is based on our recent synthesis of silica NPs with grafted low molecular weight ligands that provide steric stabilization in high salinity synthetic seawater (SSW) and API brine.[5] A major challenge is that the requirements for colloidal stability and amphiphilicity at the C-W interface are somewhat in conflict as higher hydrophilicity favors colloidal stability in brine; however, if it is too high, the HCB (amphilicity) is too much towards brine for foam stabilization. The proper balance of colloidal stability in brine and amphiphilicity for C/W foam stabilization was found for three nonionic ligands: an ether diol (GLYMO) and two short poly(ethylene glycol) ligands with 6-12 EO repeat units. Additionally, the previous synthesis[5] was modified to produce very small (ca. 7 nm diameter) GLYMO-coated NPs, whereby the viscosity is shown to be ca. 2 fold higher than for the case of larger 18 nm PEG(6-9EO)-coated NPs. For the secondary objective of this work, we identify pairs of surface-modified silica NPs and anionic, nonionic, or cationic surfactants that do not aggregate in API brine and stabilize C/W foam in porous media. The ligands on the silica surface will be shown to provide a judicious combination of sufficient $CO_2$-philicity (that is an appropriate HCB) and solvation in high salinity brine needed to stabilize C/W foam at high salinity. To our knowledge, this is the first study to report C/W foams at high salinities stabilized with nanoparticles with well defined surface coatings.

**EXPERIMENTAL**

**Materials.** NexSil 6 spherical colloidal silica nanoparticles were purchased from Nyacol Nano Technologies and received as a 17 wt.% dispersion. Silylating agents containing 6-9 EO units (2-[methoxy (polyethyleneoxy) 6-9propyl] trimethoxysilane, PEG(6-9EO), 90%, Cat. No. SIM6492.7), 8-12 EO units ([hydroxyl (polyethyleneoxy) propyl] triethoxysilane, PEG(8-12EO), 50% in ethanol, Cat. No. SIH6188.0), or sulfobetaine (3-([dimethyl (3-trimethoxysilyl) propyl] ammonio) propane-1-sulfonate, SB, 95%, Cat. No. SID4241.0) were purchased from Gelest (see Figure S1 for structures). (3-glycidyloxypropyl) trimethoxysilane (≥98%, Sigma-Aldrich), HCl (1N solution, Fisher Scientific), NaOH (1N solution, Fisher Scientific), NaCl (ACS Grade, Fisher Scientific), $CaCl_2 \cdot H2O$ (ACS Grade, Amresco), triethylene glyclol (TEG, 99%, Acros Organics), and synthetic seawater (SSW, Cat. No.8363-5, Lot 1306873, ASTM D1141, pH 8.2, Ricca Chemical Co.) were used as received. Deionized (DI) water (Nanopure II, Barnstead, Dubuque, IA) was used for all experiments. NPs coated with PEG(6-9EO) (all), PEG(8-12EO), and GLYMO (2.9) were synthesized by a previously reported method.[5] A new synthesis method to produce small (ca. 7 nm diameter) GLYMO-coated NPs is given in the next section. Surfactants were obtained from commercial sources and their structures and abbreviations are given in Table 1.

**Table 1.** Surfactants used in this study.

| Surfactant | Structure | Abbreviation |
|---|---|---|
| Sodium lauryl ether sulfate, CS-330 from Stepan Chemical Co. | | SLES |
| Sodium alpha-olefin sulfonate (12 carbons), Witconate AOS-12 from AkzoNobel | | AOS12 |
| Sodium alpha-olefin sulfonate (14-16 carbons), Witconate 96A from AkzoNobel | | AOS14-16 |
| Ethoxylated nonylphenol (30 EOs), Surfonic N-300 from Huntsman Chemical Co. | | N-300 |
| Linear alcohol ethoxylate (12-14 carbons and 22 EOs), Surfonic L24-22 from Huntsman Chemical Co. | HO$\left[\smile O\right]_{22}\left[\diagdown\right]_{5\text{-}6}$ | L24-22 |
| Linear alcohol ethoxylate (12-15 carbons and 9 EOs), Neodol 25-9 from Shell | HO$\left[\smile O\right]_{9}\left[\diagdown\right]_{10\text{-}13}$ | N25-9 |
| Dodecyltrimethylammonium bromide, Product No. D5047 from Sigma Aldrich | Br$^{-}$ N$^{+}$ $[\ ]_5$ | C12TAB |
| Hexadecyltrimethylammonium bromide, Product No. H9151 from Sigma Aldrich | Br$^{-}$ N$^{+}$ $[\ ]_7$ | C16TAB |

**Synthesis of ca. 7 nm diameter GLYMO-coated silica.** The modified synthesis of GLYMO-coated silica nanoparticles was based on that of a previous study.[5] An aliquot of (3-glycidyloxypropyl) trimethoxysilane (typically 0.5 mL for a 1 g batch of NexSil 6 particles) was added to DI water at pH 12.0 (pH adjusted by 2.5N NaOH) (typically in a ratio of 1:6 by volume) and stirred for 2 minutes at room temperature to perform base-catalyzed ring opening shown in Scheme 1 to form GLYMO.[28]

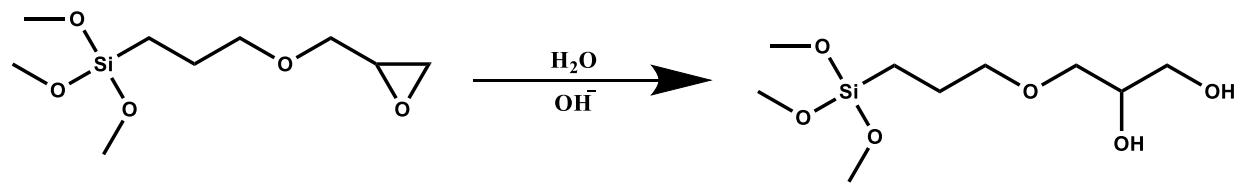

**Scheme 1.** Ring opening of (3-glycidyloxypropyl)trimethoxysilane to form GLYMO. The solution rapidly turned from turbid to clear as the reaction proceeded due to the high solubility of the GLYMO product and low solubility of the (3-glycidyloxypropyl) trimethoxysilane reactant. The resulting solution of GLYMO starting material (shown in Scheme 1 and Figure S1) was immediately added dropwise to a vial containing nanoparticles and DI water such that the final concentration of nanoparticles was 10% w/v. The amount of GLYMO added was varied from 1 to 5 µmol GLYMO/m$^2$ of silica nanoparticle surface

(determined by BET) to achieve various GLYMO coverages on the NPs. The reaction pH was maintained around 10 and the solution was stirred overnight at 40°C.

For small scale reactions (~ 10 mL, 1g Nexsil6), the resulting dispersion of grafted nanoparticles was washed five times with DI water using 30 KDa MWCO centrifuge filters at 5500 rpm for 15 minutes to remove unreacted ligands and reaction byproducts. After the final filtration, DI water was added to the retentate to bring the concentration to ca. 10% w/v silica NPs. The dispersion was bath sonicated for 15 min. and passed through a 0.45 µm syringe filter to remove any large aggregates that might have formed during the purification process.

For large scale reactions (~100 mL, 10 g Nexsil6), the samples were first centrifuged at 5,000 rpm for 10 minutes to remove large aggregates. The supernatant was collected and purified by tangential flow filtration using a Spectrum Labs KrosFlo Research II unit (SpectrumLabs, Los Angles, USA) with 30 KDa MWCO PES hollow fibers. The sample was first diafiltered with DI water (approx. 5 times the original sample volume) followed by concentration to the original volume (in this case ~100 mL). The dispersion was centrifuged to remove any larger NPs and then bath sonicated for 20 minutes.

An aliquot of the dispersion was dried overnight at 80°C to determine the final concentration of silica + GLYMO coating in the dispersion by mass. The NP concentrations in this manuscript are given in terms of total mass of solids in the dispersion (i.e. total mass of silica particles + grafted ligands).

**Air-water surface tension measurements.** The surface tension of NP and surfactant solutions was determined with axisymmetric drop shape analysis of a pendant droplet formed on a stainless steel capillary. The apparatus was a Theta Optical Tensiometer (Biolin Scientific). Droplets were equilibrated for 2-10 minutes until a steady-state surface tension value was reached. Recorded images of the bubbles were analyzed using the OneAttension software package (Biolin Scientific). The mean and standard deviation of ca. 200 measurements taken over 10 s were recorded.

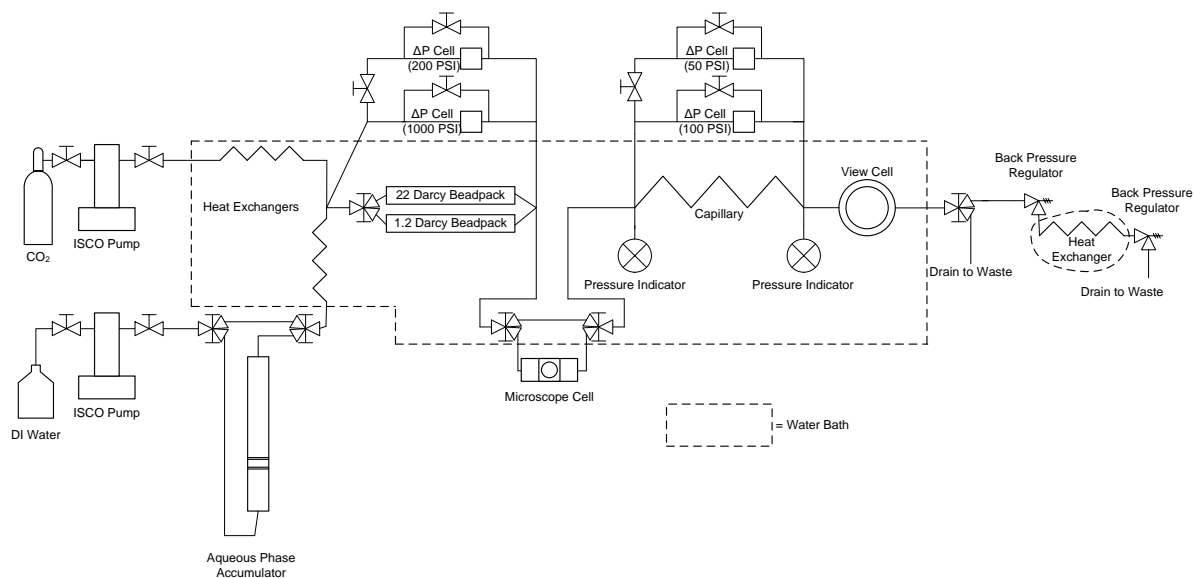

**Figure 1.** Apparatus for C/W foam generation, measurement of viscosity, and observation of foam stability.

**C/W foam formation and apparent viscosity measurement.** C/W foams were formed and characterized in an apparatus described elsewhere,[8, 29-32] as shown in Figure 1. All experiments were done at 2800 psia ±100 psi and RT or 50°C, as indicated. The pressure gradient caused by the flow of foam through the beadpack or capillary tube was measured with differential pressure transducers. The beadpack used in this study had a permeability of 22 d (0.38 cm ID x 11.3 cm long), filled with 180 µm spherical glass beads (porosity of 0.34, pore volume of 0.436 mL). In the 22 d beadpack, a flow rate of 3 mL/min gives a superficial velocity of 1250 ft/day, a shear rate of 2270 $s^{-1}$, and a residence time of 9 s. In the capillary tube (0.0762 cm ID x 195 cm long) at a flow rate of 3 mL/min, the velocity was 31000 ft/day, the shear rate was 1150 $s^{-1}$, and the residence time was 18 s. The apparent viscosity of the foam in the beadpack or capillary tube was calculated using Darcy's Law or the Hagen-Poiseuille equation, respectively.[24] Per the manufacturer's information, the accuracy of the differential pressure reading is ± 0.25% of full scale.

**Determination of NP stability with surfactants at ambient conditions.** Nanoparticle dispersions at a concentration of 0.5% w/v were made in either (1) DI or (2) NaCl and/or $CaCl_2$ solutions to a final salt concentration of either 4% NaCl or 8% NaCl + 2% $CaCl_2$ (API brine). The pH was left unadjusted (pH~8.5) or adjusted to 3.5 with 1N HCl and measured with a Mettler-Toledo FiveGo pH meter equipped with a micro pH probe. Samples were then stored at room temperature (23±1°C, RT) for 7 days and visual observations of nanoparticle dispersions were recorded.

## RESULTS AND DISCUSSION

**Properties of coated NPs.** Spherical silica NPs (NexSil 6, 7.4 nm diameter by DLS as reported previously[5] with various ligands (see Figure S1 for structures) grafted to their surfaces are presented in Table 2. For simplicity, the NP-ligand complex is referred to by the ligand used since the core NP was the same in all cases. When multiple ligand coverages were investigated, the coverage is denoted in parentheses in $\mu mol/m^2$, for example "GLYMO (1.2)" indicates GLYMO-coated silica NPs with ligand coverage of 1.2 $\mu mol/m^2$ as determined by TGA, as discussed in further detail previously.[5] As discussed in the Experimental section, the three lowest GLYMO ligand coverages (1.2-1.8 $\mu mol/m^2$) were synthesized at 40°C after a base-catalyzed GLYMO ring-opening step. The improved synthesis gave a smaller final NP size because it reduced aggregation during the synthesis due to the lower reaction temperature and absence of NaCl produced during the previously reported synthesis.[5] In contrast, the remainder of the NPs given in Table 2 were synthesized by a recent grafting method at 60°C[5] that produced aggregates of approximately 16~21 nm diameter. As shown in Table 2, the particle organic fraction and correspondingly, ligand coverage, for a given initial amount of GLYMO added to the reaction, was slightly lower at the lower T of 40ºC.[5] However, the ligand coverage and NP size were highly controllable at a given temperature for each method. For silica NPs coated with each of these ligands, the NPs remained negatively-charged, as expected from the ungrafted SiO⁻ sites on the surfaces (bare NexSil 6 had a zeta potential of approximately -38 mV at pH~9).[5] For the four cases where the zeta potential was measured at a pH range from 2~9, the zeta potentials for all the cases at pH~9 range from 30~41 mV (Figure S4).[5]

**Table 2.** Properties of coated NPs. Core particle is 7.4 nm diameter NexSil 6 bare colloidal silica in all cases.

| Grafted ligand and coverage ($\mu mol/m^2$) | Particle organic fraction by TGA (wt.%) | Ligand coverage ($\mu mol/m^2$) | DLS size in DI @ RT (nm) | CFT in API brine, pH ~8.5, 0.5% NPs (°C) |
|---|---|---|---|---|
| GLYMO (1.2) | 6.0 | 1.2 | 8.6±0.25 | >95 |
| GLYMO (1.4) | 6.9 | 1.4 | 9.6±1.1 | >95 |
| GLYMO (1.8) | 9.1 | 1.8 | 10.1±3.1 | >95 |
| GLYMO (2.6) | 12.6 | 2.6 | 7.4±0.12 | >95 |
| GLYMO (2.9) | 13.8 | 2.9 | 16.5±2.3 | >95 |
| SB | 12.1 | 1.6 | 20.9±2.2 | >95 |
| PEG(6-9EO) (1.7) | 21.5 | 1.7 | 20.0±3.0 | 58* |
| PEG(6-9EO) (2.1) | 26.0 | 2.1 | 18.0±2.7 | 49* |
| PEG(8-12EO) | 15.3 | 0.9 | 19.6±1.2 | 50 |

* re-disperses when cooled

All of the coatings in Table 2 have been shown to stabilize NPs in API brine for up to 30 days at room temperature.[5] In agreement with previous results,[5] the visual critical flocculation temperature (CFT) for NPs with PEG-based coatings in API brine was 49-58°C, above which the NPs rapidly aggregated. The GLYMO-coated NPs did not immediately flocculate upon heating to 95°C in pH~8.5 API brine. However after ~10 hours, the dispersion of GLYMO particles with 1.2 µmol/m$^2$ ligand coverage became highly turbid and settled at 50°C in API brine at pH~8.5. For higher GLYMO coverage the aggregation and settling became apparent in 2-4 days with 1.4µmol/m$^2$, 7-14 days with 1.8 µmol/m$^2$, and more than 30 days for 2.9 µmol/m$^2$ coverage. The rapid aggregation of PEG-based NPs at their CFTs and the slow aggregation of GLYMO-coated NPs at 50ºC both serve to limit the useful T range over which these NPs are appropriate for foam generation in porous media, given that aggregation of NPs would cause blocking of porous media. As a result, all C/W foam experiments were carried out at T conditions where the NPs were stable for at least 24 h to minimize NP aggregation during the ca. 2 min. timeframe when the NPs were exposed to elevated T inside the flow apparatus.

**Table 3.** Surface tension reduction with NPs in API brine, pH~8.5.

| Grafted ligand | Surface tension ± Std. dev. (mN/m) |
|---|---|
| No NPs | 75.2 ± 0.5 |
| 1% GLYMO (2.9) | 71.3 ± 0.3 |
| 1% GLYMO (1.8) | 65.6 ± 0.5 |
| 1% GLYMO (1.2) | 58.4 ± 0.5 |
| 1% PEG(6-9EO) (1.7) | 54.4 ± 0.7 |

**Surface tension reduction with NPs.** To characterize the interfacial activity of the surface modified NPs in API brine, air-water surface tension measurements were made for the NP dispersions (Table 3). Without added NPs, the surface tension of the API brine was 75.16 mN/m, higher than that of pure water (ca. 72.8 mN/m) because of organization of the added salt ions at the air-water interface. The effect of each ion is expected to be approximately in the same order as the Hofmeister series.[33] The surface tension of a 1% dispersion of bare silica starting material was not tested because it was not stable in API brine, but the addition of hydrophilic unmodified silica NPs are not expected to change the surface tension as seen in previous studied of silica nanoparticles in water.[14, 34-35] Adding 1% GLYMO (2.9) lowered the surface tension by ca. 4 mN/m (i.e. surface pressure of ca. 4 mN/m), indicating that the large and highly hydrophilic NPs were relatively inefficient at reducing surface tension. The surface pressure was significantly higher for the smaller 7 nm NPs with the smaller levels of GLYMO with coverage up to 1.8 µmol/m$^2$. Assuming a full monolayer of grafted ligands is 7.6 µmol/m$^2$, the GLYMO coverage of 1.8 µmol/m$^2$ corresponds to 24% of a monolayer.[5] Furthermore, the surface tension decreased monotonically as the GLYMO ligand coverage was lowered reaching a quite large surface pressure of nearly 17 mN/m for GLYMO (1.2). We hypothesize that the hydrophobic region (inside the diol tip) of GLYMO is oriented towards air at the at the air-water interface, given the facile bending of the flexible ether bond, particularly when the ligands are not too crowded on the NP surface. However, the surface pressure decreases when the grafting density is too high, whereby the highly hydrophilic diol tips on the more extended chains are preferentially forced to the surface. Here, the more hydrophilic tips lower the adsorption at the air-water interface.

The longer ligand PEG(6-9EO) with numerous flexible ether bonds was the most efficient at reducing IFT with a corresponding surface pressure of ca. 20 mN/m at a coverage of 1.7 µmol/m$^2$. A similar surface pressure of ca. 20 mN/m was reported in DI water by Stocco et al. for fumed silica NPs with highly hydrophobic dimethyl ligands grafted on the surfaces.[14] Silica nanoparticles with short-chain PEG ligands are known to be interfacially active in $CO_2$-water systems[2] and NPs with longer PEG-based polymers have been demonstrated to efficiently reduce air-water surface tension[19] and oil-water interfacial tension.[23]

The adsorption energy of a spherical particle is a function of the loss of air-water interface of high tension, which is replaced by air-NP and water-NP hemispherical interfaces. Given that the contact angle of the NP was unknown, a value of 90° was used as in previous studies.[36] For a spherical particle with an air/water/NP contact angle of 90° and a measured surface pressure ($\gamma_o - \gamma$),

$$\Delta E = \frac{-(\gamma - \gamma_o)\pi a^2}{\eta} \qquad (1)$$

where $a$ is the particle radius and $\eta$ is the 2-dimensional packing fraction. For simplicity, we assume a close-packed interface where $\eta = 0.91$. Nanoparticles with $a$ on the order of 10 nm are known to adsorb essentially irreversibly at air-water[17, 19, 21] and oil-water[21] interfaces. For example, small NPs (radius of ca. 10 nm) of IO with adsorbed 2-(2-methoxyethoxy) ethyl methacrylate MEO$_2$MA homopolymer[17] or MEO$_2$MA- oligo(ethylene glycol) methacrylate (OEGMA) copolymer,[19] or silica with grafted poly(2-(dimethylamino)ethyl methacrylate) (PDMAEMA) brushes[21] at air-water interfaces had a calculated $\Delta E$ of ca. -10$^3$ kT at room temperature. For a spherical GLYMO (1.2)-coated silica particle with radius of 4.3 nm adsorbed at an air-API brine interface, the calculated $\Delta E$ values are of the order -10$^2$ kT. For the larger ca. 11 nm radius PEG(6-9EO) (1.7)-coated silica particles, the calculated $\Delta E$ values are of the order -10$^3$ kT. Continuing this trend, fumed silica NPs with a radius of 50 nm were estimated to have an adsorption energy on the order of -10$^4$ kT at a $CO_2$-water interface using a calculation method based on an estimate of interfacial area covered by one particle,[2] rather than measuring surface pressure as used in Eqn. 1. The essentially irreversible adsorption of a NP at a fluid-fluid interface may be contrasted with the much more dynamic adsorption and desorption of free surfactants, where $\Delta E$ values are of the order -10$^0$ kT.[1] Together with literature data available on similar surface-modified NPs,[2] the surface tension results presented in Table 3 suggest that GLYMO (1.8)-, GLYMO (1.2)-, and PEG-coated NPs could potentially be efficient C/W foam stabilizers, as $CO_2$-water surface pressures are somewhat related to air-water surface pressures.[37]

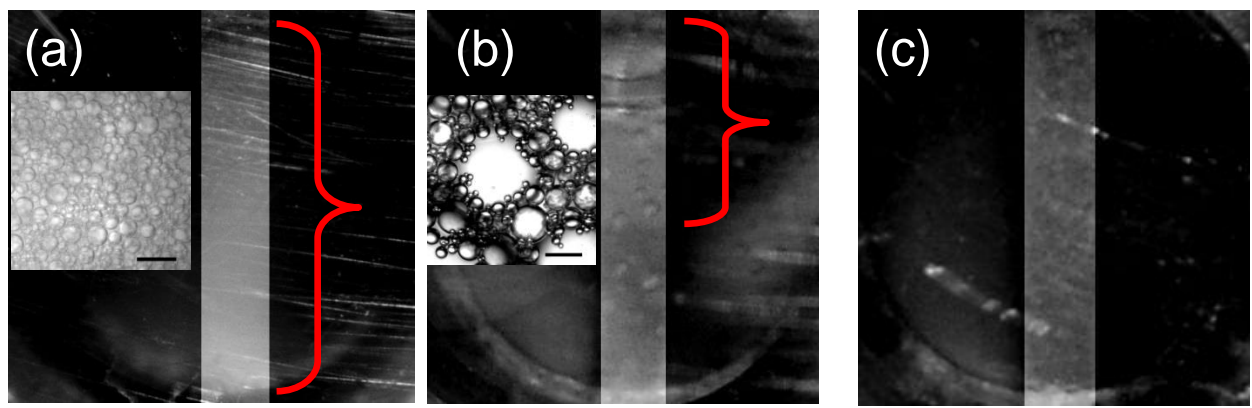

**Figure 2.** Representative examples of C/W foam in the view cell at 2800 psia with (a) fine texture (inset shows optical micrograph of finely-textured foam, scale bar = 100 µm), (b) intermediate texture (inset shows optical micrograph of coarsely textured foam, scale bar = 100 µm), and (c) no foam. Foam is visible in the center of the window in the vertical channel between Teflon spacers and the approximate foam height is shown with red brackets. The Teflon spacers have been darkened with photo editing software to highlight the channel. Dark regions in the channel indicate absence of foam. Window diameter is 1.4 cm and visual path length is 0.8 cm.

**C/W foams stabilized by NPs only.** C/W foams were generated in a 22 d beadpack at 3 mL/min of total fluid flow and a pressure of 2800 psia. Residence times, shear rates, and other flow properties are given in the Experimental section. Representative examples of foams observed in a high pressure microscopic view cell downstream of the beadpack foam generator and in a macroscopic view cell downstream of the capillary tube are given in Figure 2. Highly viscous foams with apparent viscosities of >10 cP in the capillary tube filled the view cell with an opaque white appearance and bubble sizes of < 50 µm diameter, as shown in Figure 2a. Foams with apparent viscosities between ca. 2 cP and 10 cP in the capillary tube typically did not fill the view cell completely, as shown in Figure 2b. In Figure 2b, the bottom part of the macroscopic view cell is filled with excess aqueous phase which was not expelled from the view cell due to the low viscosity of the foam. Cases where no significant foam was formed are shown in Figure 2c, where the view cell is filled with transparent aqueous phase and $CO_2$.

Several trends are observable in the apparent viscosity of NP-stabilized foams measured by pressure drops in both the beadpack and the capillary tube are given in Table 4. For GLYMO (1.4) NPs in DI water, no foam was observed, but increasingly viscous foams were produced as the salinity was increased from that of SSW to API brine at RT and 75% quality. When the T was increased to 35°C, there was a slight decrease in foam viscosity in both the beadpack and capillary tube. Higher temperatures were not investigated with GLYMO (1.4)-coated NPs because they slowly aggregated as discussed above. With GLYMO (1.8)-coated NPs, however, temperatures of both 35°C and 50°C were investigated given the improved colloidal stability at this higher GLYMO coverage. The decrease in beadpack foam viscosity was only slight at 35°C compared to RT (from 42.0 cP to 34.3 cP), but was more significant at 50°C falling to 3.5 cP. With a higher GLYMO coverage of 2.9 µmol/m$^2$, an apparent viscosity of less than 2 cP was observed at all conditions tested, 20 fold less than for the GLYMO (1.8)-coated NPs at RT and 75% quality. The low viscosity may be expected from the lower surface pressure for the air-brine interface (high hydrophilicity and hydrophilic-$CO_2$ philic balance, HCB) and the large NP size. Larger NPs are less efficient in stabilizing the interface given a lower NP number concentration per mass and greater distances of NP mass from the interface. Similarly, the highly hydrophilic SB-coated NPs did not stabilize viscous foam at the conditions tested. PEG(6-9EO) (1.7)-coated NPs stabilized foam up to 43°C, below the NP CFT of 58°C (Table 1). C/W foam formation was not investigated above the CFT due to potential plugging of the porous media with large NP aggregates. In the beadpack, PEG(6-9EO) (2.1)-coated NPs generated foams in SSW with an apparent viscosity of 6.6 cP and an even higher value of 13.0 cP in API brine at RT and 75% quality. At a higher quality of 95%, foam was not formed in API brine. When the temperature was increased to 40°C, approaching the CFT of 49°C (Table 1), weak foam was formed. Foam was also generated with PEG(8-12EO)-coated NPs in SSW with a low viscosity that increased modestly upon changing to API brine at RT and 75% quality.

**Table 4.** Viscosity of NP-only foams formed in 22 d beadpack at 3mL/min at 2800 psia. NP concentration was 1%. "No foam" indicates no foam was observed in the macroscopic view cell and viscosities were < ca. 2 cP.

| Ligand on NPs | NP diameter by DLS (nm) | Salinity | T, °C | Quality | µ$_{app}$, beadpack, cP | µ$_{app}$, capillary, cP |
|---|---|---|---|---|---|---|
| GLYMO (1.4) | 9.6±1.1 | DI | RT | 75% | No foam | No foam |
| GLYMO (1.4) | 9.6±1.1 | SSW | RT | 75% | 25.7 | 13.8 |
| GLYMO (1.4) | 9.6±1.1 | API | RT | 75% | 34.5 | 33.0 |
| GLYMO (1.4) | 9.6±1.1 | API | 35 | 75% | 29.8 | 29.0 |
| GLYMO (1.8) | 10.1±3.1 | API | RT | 75% | 42.0 | 40.6 |
| GLYMO (1.8) | 10.1±3.1 | API | 35 | 75% | 34.3 | 36.8 |
| GLYMO (1.8) | 10.1±3.1 | API | 50 | 75% | 3.5 | 7.9 |
| GLYMO (2.6) | 7.4±0.12 | API | RT | 75% | 26.12 | 26.91 |
| GLYMO (2.9) | 16.5±2.3 | API | RT | 75% | No foam | No foam |

| GLYMO (2.9) | 16.5±2.3 | API | RT | 95% | No foam | No foam |
| GLYMO (2.9) | 16.5±2.3 | API | 50 | 95% | No foam | No foam |
| SB | 20.9±2.2 | API | RT | 75% | No foam | No foam |
| PEG(6-9EO) (1.7) | 20.0±3.0 | API | 43 | 75% | 9.3 | 16.0 |
| PEG(6-9EO) (2.1) | 18.0±2.7 | SSW | RT | 75% | 6.6 | 0.8 |
| PEG(6-9EO) (2.1) | 18.0±2.7 | API | RT | 75% | 13.0 | 10.0 |
| PEG(6-9EO) (2.1) | 18.0±2.7 | API | RT | 95% | No foam | No foam |
| PEG(6-9EO) (2.1) | 18.0±2.7 | API | 40 | 75% | 3.1 | 2.8 |
| PEG(8-12EO) | 19.6±1.2 | SSW | RT | 75% | 4.3 | 3.8 |
| PEG(8-12EO) | 19.6±1.2 | API | RT | 75% | 7.3 | 7.0 |

The data presented in Table 4 suggest that both GLYMO- and PEG-based coatings are sufficiently $CO_2$-philic to adsorb at the $CO_2$-water interface to stabilize foams. The small, moderately-coated GLYMO (1.4)-coated NPs formed viscous C/W foams only in high salinity brine, indicating that they were too hydrophilic in DI water and/or were repelled from the interface due to strong NP-NP and NP-interface electrostatic repulsion.[38] We note that the apparent viscosities of 42.0 cP in the beadpack and 40.6 cP in the capillary tube measured with GLYMO (1.8)-coated NPs in API brine at RT and 75% quality are the highest reported C/W foam viscosities for a NP stabilized foam generated in porous media (without additional free surfactant or polymers).[2, 39-40] We hypothesize that the small size of the GLYMO (1.8)-coated NPs may have significantly improved foam viscosity in porous media because of: (1) more rapid diffusion to the interface of smaller NPs and (2) higher number concentration per given mass concentration. A larger amount of the interface may be covered with the higher number concentration with more of the particle mass close to the interface. For all four NPs tested at elevated T, an increase in T decreased the foam viscosity compared to the RT experiment. We hypothesize that the reduction in foam apparent viscosity is due to an increased rate of coalescence due to more rapid lamella drainage and hole formation[41] and a decrease in the viscosity of the aqueous lamellar phase.[42] In the case of C/W foams stabilized by nonionic surfactants presented by Adkins et al.,[41] foam viscosity decreased as temperature increased, and no foams were formed more than ca. 5°C above the cloud point of the surfactant. In the present study, a similar behavior was observed for PEG-coated NPs in that foam viscosities decreased as the CFT was approached and the ligands became insufficiently solvated.

**New NP+surfactant formulations for C/W foams.**

Interactions between surface-modified silica nanoparticles and surfactants have a large effect on the properties of emulsions and foams relevant to a range of applications. An understanding of the phase behavior is imperative for designing successful stabilizers for foam generation in porous media. Table 5 summarizes visual observations of a wide range of dispersions composed of NPs and surfactants. To mimic the conditions that the particles encounter during a $CO_2$ foam formation experiment, two pH's were chosen: the unadjusted pH of the NP dispersion (pH~8.5) expected at ambient conditions and a low pH of 3.5 to mimic a $CO_2$-saturated aqueous phase.

Bare NexSil 6 particles were chosen as a reference or control. They became unstable when mixed with nonionic or cationic surfactants even in DI water. As described in Table S1, H-bonding or electrostatic attraction between nonionic or cationic surfactants, respectively, and anionic silica NPs commonly causes aggregation of the NPs at low salinity. In 4% NaCl or API brine, bare particles are unstable regardless of surfactant choice or pH due to the strong screening of electrostatic repulsion between the particles.[5] In contrast, the findings in Table 5 indicate that bare anionic silica NPs do not strongly attract anionic surfactants (as summarized in Table S1), as the mixtures were stable in DI water. However, both the nonionic and cationic surfactants adsorbed on the NP surfaces and caused flocculation in DI water. Given that so few of these mixtures were stable, and none in brine, an alternative strategy was chosen to design mixtures of surface modified NPs with surfactants.

**Table 5.** Summary of stability observations of 1% NPs in 0.5% surfactant solutions at room temperature and ambient pressure.

| NP | SLES | | AOS12 | | AOS14-16 | | N-300 | | L24-22 | | C12TAB | | C16TAB | |
|---|---|---|---|---|---|---|---|---|---|---|---|---|---|---|
| | DI | API | DI | API | DI | 4% NaCl | DI | API | DI | API | DI | API | DI | API |
| Bare NexSil 6 | ✓ | XX | ✓ | XX | ✓ | X | XX | XX | XX | XX | XX | XX | XX | XX |
| NexSil 6 + | ✓ | ✓† | ✓ | ✓† | ✓ | ✓ | ✓ | ✓† | ✓ | ✓† | * | ✓† | * | ✓† |

| GLYMO (2.9) | | | | | | | | | | | | | | |
|---|---|---|---|---|---|---|---|---|---|---|---|---|---|---|
| NexSil 6 + SB | ✓ | ✓† | ✓ | ✓ | ✓ | ✓† | ✓ | ✓ | ✓ | ✓ | ✓ | ✓ | * | ✓ |
| NexSil 6 + PEG (6-9EO) (2.1) | ✓ | ✓ | ✓ | ✓† | ✓ | ✓ | ✓ | ✓ | ✓ | ✓ | ✓ | ✓† | ✓ | ✓† |

✓ = no observable change in 7 days at pH 3.5
XX = rapid aggregation and settling of particles at pH~8.5, do not redisperse when pH is reduced to 3.5
X = slow aggregation of particles over 1-2 days at pH 3.5
* = rapid aggregation of particles at pH~8.5, redispersion occurs when pH is reduced to 3.5 and the NPs then remain unchanged for 7 days
† = NP+surfactant combination was also used for C/W foam formation (See Tables S3 and S4 for these and other examples)

It is well known that grafting of ligands[5] and polymers[43-45] to NPs can provide steric repulsion to prevent aggregation of NPs in brine at conditions where electrostatic repulsion is weakened by charge screening. Furthermore, Table 5 shows that grafted nonionic or zwitterionic ligands can also prevent NP aggregation in the presence of various nonionic or cationic surfactants by mitigating the NP-surfactant attraction. Each of these surface modifications prevented aggregation, presumably by due to interactions/adsorption of nonionic surfactants N-300 and L24-22 at both salinities and pH values. The stability may possibly be due to disruption of adsorption of the bulky EO surfactant headgroups on the NP surface (see Table 2 for surfactant structures). Furthermore, the removal of anionic $SiO^-$ sites via grafting lowers the propensity for $Ca^{2+}$ bridging. However, differences in colloidal behavior were apparent upon adding cationic surfactants that interact strongly with unreacted $SiO^-$ groups and consequently add hydrophobic surfactant tails to the surface. The quaternary amine cationic surfactants also have a much smaller headgroup than the nonionic surfactants investigated, which raises the tendency for adsorption on the NP surfaces between grafted ligands.

For GLYMO (2.9)-coated NPs, rapid aggregation only occurred with the cationic surfactants C12TAB and C16TAB at unadjusted pH. Interestingly, when the pH was reduced to 3.5, the sample with C12TAB in DI water redispersed to form a stable dispersion. The surfactant charge is positive at each pH, but the NP surface becomes much less negatively charged as the silanol groups are more protonated at low pH.[5] By reducing the electrostatic interaction upon neutralizing SiO$^-$ sites on the NP surface upon protonation, we hypothesize that the surfactant desorbs, the NPs become more hydrophilic and redisperse, as observed experimentally. In contrast, the SB-coated NPs interacted more weakly and only became unstable at their unadjusted pH with the most hydrophobic cationic surfactant C16TAB. Finally, aggregation was not observed with the longest ligand, 6-9EO PEG, even though it was present at the lowest coverage on any of the NPs (Table1). We note that only the 6-9EO PEG ligand is longer than the C16TAB surfactant, and thus could provide steric repulsion against particle aggregation even with surfactant adsorbed on the core particle surface.

The discovery of ligands that provide colloidal stability in the presence of concentrated brines (especially API brine)[5] and surfactants (Table 5) provided a basis for the design of many new NP + surfactant formulations for reduction of interfacial tension (Figure S3 and Table S2) and generation of C/W foam in porous (see Tables S3 and S4). When 1% GLYMO (2.9) was added to AOS12 and 1% PEG(6-9EO) (1.7) to N25-9, the $CO_2$-API brine interfacial tension did not decrease(Figure S3). When interfacially active PEG(6-9EO) (1.7)-coated NPs were used with a low concentration (0.0001%) N25-9 surfactant, the NPs lowered the surface tension below that of the surfactant-only value.  However, when higher surfactant concentrations were used, the mixture gave the same surface tension values as the surfactant-only case. Perhaps the lack of an effect for GLYMO (2.9) is a consequence of the small reduction in air-brine surface tension in Table 3.  In the future it would be warranted to study other combinations of the NPs and surfactant to continue to look for synergies in lowering the interfacial tension.

All surfactants tested could stabilize C/W foam without added NPs. Remarkably, many combinations of surfactants with the surface functionalized NPs formed viscous stable foams at salinities up to API brine and temperatures up to 50°C as presented in Tables S3 and S4 and described in detail in the supporting information section. In the remainder of this section, we give only a summary of these results.. In API brine (Table S3), addition of weakly interfacially active GLYMO (2.9)-coated NPs (on the basis of the interfacial tension measurements above) did not significantly affect the viscosity and stability of the C/W foams relative to the case of surfactant only.  This lack of an effect is attributable in part to the low interfacial activity of the NPs without surfactant. Furthermore, the NP charge may limit the kinetics and thermodynamics of NP adsorption at the interface, particularly for the anionic and nonionic surfactants.  When strongly interfacially active NPs such as those with PEG-based coatings were added, the C/W foam viscosity tended to decrease compared to the case of surfactant only, suggesting the unadsorbed NPs may have adsorbed surfactant that otherwise would have been available to adsorb at the $CO_2$-water interface (Table S3). In terms of foam stability via bubble size measurement (Table S4), addition of 1% slightly-interfacially active GLYMO (2.9)-coated NPs did not significantly improve the stability of the C/W foams, with the exception of one point collected with 0.5% N-300 in API brine at RT and 95% quality. For this case, the stability was improved by an order of magnitude (Table S4). In a second example, when 1% PEG(6-9EO) (2.1)-coated NPs were added to 0.5% C16TAB surfactant in API brine at RT and 95% quality, the foam stability improved by an order of magnitude to equal that of the stability observed with NPs only (Table S4).

**CONCLUSIONS**

GLYMO, PEG(6-9EO), and PEG(8-12EO) ligands on the surface were sufficiently $CO_2$-philic to tune the hydrophilic/$CO_2$-philic balance of silica NPs to allow stabilization of both bulk C/W foam and C/W foam in a glass beadpack. Thus, foam formation did not require the use of any free (i.e. ungrafted) surfactant. An improved grafting process, relative to our previous study,[5] was used to produce small, surface-modified primary silica NPs through use of a lower temperature and prevention of NaCl formation during the synthesis. The resulting ca. 10 nm diameter GLYMO-coated silica NPs stabilized foams with apparent viscosities of over 40 cP in both beadpack and capillary tube in API brine at RT. Larger (ca. 16.5 nm diameter) GLYMO (2.9)-coated NPs synthesized by the previously-reported technique[5] did not stabilize foam of over 2 cP apparent viscosity at any condition tested because of the large size and high hydrophilicity of the GLYMO coating at a high graft density. In contrast, ca. 18 nm diameter PEG(6-9EO)-coated NPs produced C/W foams with lower viscosities up to 13.0 cP in the beadpack at RT and 16.0 cP in the capillary tube at 43°C due to the amphiphilic PEG-based coating. Additionally, NPs with three ligands, which were previously demonstrated to be sufficiently solvated in high salinity API brine to provide long-term colloidal stability,[5] are shown in the present study to form stable dispersions with various free molecular surfactants in DI water and even API brine at RT. A wide variety of NPs + anionic, nonionic, or cationic surfactant mixtures that formed stable dispersions were also found to stabilize C/W foams in porous media at high salinity. The apparent viscosities were generally comparable to (or less than) the values for the surfactant without the presence of added NPs. In two cases, however, an order of magnitude increase in C/W foam stability was observed upon adding NPs. Colloidal stability of the NPs (with or without free surfactant) prevented aggregation of either the NPs or surfactants in the aqueous lamellae, which would otherwise destabilize the foam and/or block pores in the glass beadpack. With the remarkable combination of sufficient $CO_2$-philicity (appropriate hydrophilic-$CO_2$ balance) and solvation in high salinity brine provided by the grafted ligands, these functionalized NPs are the first examples with well-defined surface coatings to stabilize C/W foam at high salinity. These results provide a basis for future studies of the mechanism of foam stabilization by NPs and NP/surfactant mixtures at high salinity.

## ASSOCIATED CONTENT

**Supporting Information**
Expected interactions of NP-surfactant combinations, structures of ligands, high pressure interfacial tensiometer, $CO_2$-water interfacial tension measurements, viscosity summary of NP + surfactant foams formed in 22D beadpack at 3mL/min at 2800 psia, zeta potential measurement, C/W foam stability determination, and NP+surfactant formulations for C/W foams.

## AUTHOR INFORMATION

**Notes**
The authors declare no competing financial interest.

## ACKNOWLEDGMENT

This work was partially supported by the Gulf of Mexico Research Initiative and the DOE center for Frontiers of Subsurface Energy Security. KPJ and TMT also acknowledge the Robert A. Welch Foundation (F-1319 and F-1696, respectively). The authors thank Hugo Celio for assistance with TGA.

**Supporting information**

Table S1. Expected interactions of NP-surfactant combinations

| | | Particle charge | | |
|---|---|---|---|---|
| | | + (e.g. alumina NPs at neutral pH) | - (e.g. bare colloidal silica NPs at neutral pH) | "weak" - (e.g. silica NPs with grafted surface modifiers) |
| Surfactant charge | + (e.g. quaternary amines and protonated primary, secondary, and tertiary amines) | *Electrostatic repulsion* | *Electrostatic attraction*<br><br>Reduced surface charge of particles below ~CMC, particle aggregation at low concentrations of surfactant, reduced emulsion droplet diameter than surf. only, increased emulsion volume than surf. only, increased IFT due to competition for surfactant [26]<br><br>Increased IFT due to competition for surfactant and increased dilational viscoelasticity of interface [34]<br><br>Attraction keeps particles in thinning films for improved stability [46] | *Electrostatic attraction, maybe H-bonding (depending on structures)*<br><br>Possible loss of colloidal stability at low surf. concentrations, depending on particle coating properties |
| | "weak" + (e.g. protonated carboxy betaines) | *Electrostatic repulsion* | *Electrostatic attraction*<br><br>Reduced surface charge of particles, may cause aggregation of particles, decreased droplet size compared to surf. only, increased emulsion volume compared to surf. only [24] | *Electrostatic attraction, maybe H-bonding (depending on structures)*<br><br>Possible loss of colloidal stability in brine, depending on surfactant tail length and particle coating properties |
| | - (e.g. sulfonates and deprotonated | *Electrostatic attraction*<br><br>With poor surfactants (short | *Electrostatic repulsion*<br><br>Repulsion drives particles from thinning films [46] | *Electrostatic repulsion*<br><br>Stable when both NPs and |

| | | | |
|---|---|---|---|
| | carboxylic acids) | carboxylic acids): Reduced surface charge of particles, increased emulsion volume than surf. only, increased droplet stability than surf. only, *decreased* IFT through cooperative adsorption at interface [47] | Via simulation: Decrease in IFT compared to surf. alone due to repulsive interactions driving surfactant to fluid-fluid interface [48]<br><br>Singh paper and Rouhi's paper are the only examples I know of where -/- worked well together. | surfactants were independently stable |
| | none (e.g. ethoxylated nonionics) | *vdW attraction, maybe H-bonding (depending on structures)* | *vdW attraction, maybe H-bonding (depending on structures)*<br><br>Increased hydrophobicity of particles below ~CMC, particle aggregation over a region of [surf.] possibly due to "micelle bridging", increased emulsion volume than surf. only, increased droplet stability than surf. only, increased IFT due to competition for surfactant [25]<br><br>Increased emulsion stability when particles were aggregated, increased IFT due to competition for surfactant [49] | *vdW attraction, maybe H-bonding (depending on structures)*<br><br>Possible loss of colloidal stability, depending on particle coating properties |
| | Zwitterionic[50] (e.g. sulfobetaine) | *Electrostatic attraction from - group in surf.* | *Electrostatic attraction from + group in surf.*<br><br>Weaker attraction than for purely + surfactants, adsorption extent increases with tail length [51]<br><br>Very weak attraction when surf. has short tail [52] | *Electrostatic attraction from + group in surf., maybe H-bonding (depending on structures)*<br><br>Possible loss of colloidal stability, depending on particle coating properties |

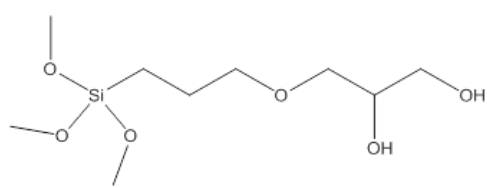 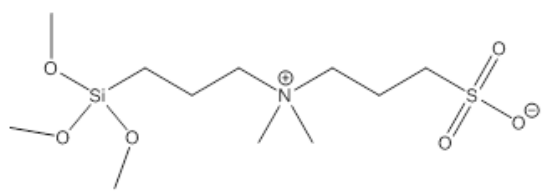

GLYMO                                   SB

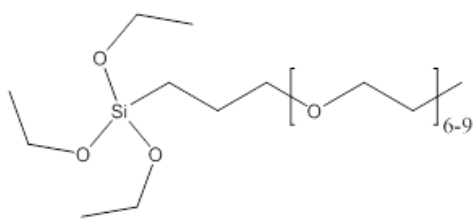 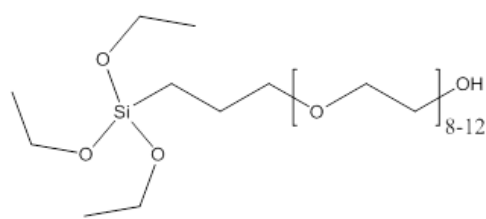

PEG(6-9EO)                              PEG(8-12EO)

Figure S1. Structures of ligands.

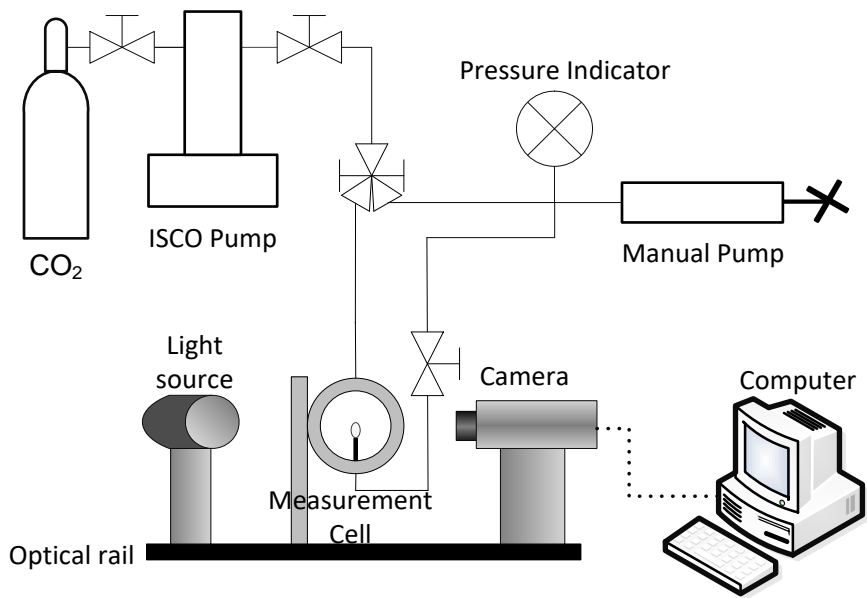

Figure S2. High pressure interfacial tensiometer.

**CO$_2$-water interfacial tension measurements**

The interfacial tension between CO$_2$ and aqueous phases containing surfactant and/or nanoparticles was determined with axisymmetric drop shape analysis of a pendant CO$_2$ bubble, where equipment and techniques were adapted from previous studies.[8, 37, 53] As shown in Figure S2, the apparatus consisted of a 28 mL stainless steel variable-volume view cell ("Measurement cell") containing a movable piston, two syringe pumps, an optical rail for alignment, a monochromatic light source, a video camera, and a computer (Figure S2). Pendant CO$_2$ bubbles were formed at the end of a PEEK capillary (0.057 in. O.D x 0.01 in. I.D.) in the variable-volume view cell containing 10~16 mL of aqueous phase with a known concentration of surfactant and/or nanoparticles. Prior to the formation of pendant bubbles, the aqueous phase was stirred for 20 min in the presence of excess CO$_2$ to saturate the aqueous phase. The typical size of the captive bubble was 2–4 mm in diameter. Bubbles were equilibrated for at least 400s prior to imaging. Recorded images of the bubbles were analyzed using a software package (CAM200, KSV Ltd., Finland) to estimate the interfacial tension. The mean and standard deviation of 10-20 measurements taken at least 10 s apart were recorded. The pressure in the cell was controlled with an ISCO syringe pump (Teledyne ISCO, Lincoln, NE) in constant pressure mode connected to the measurement cell. For safety, the measurement cell was located behind a 3/8 in. thick polycarbonate shield.

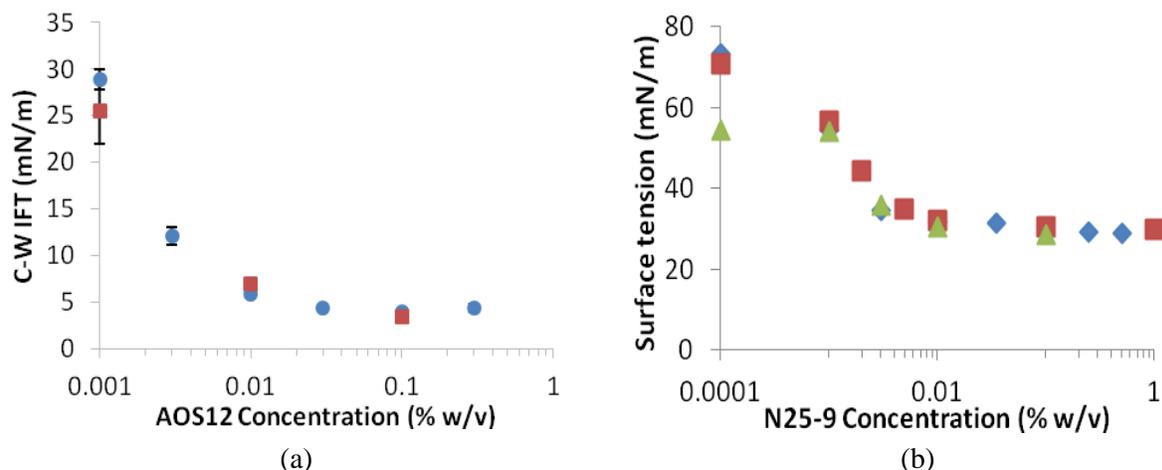

(a)                                          (b)

Figure S3. (a) $CO_2$-API brine interfacial tension at RT and 2800 psia of AOS12 with (squares) and without (circles) 1% GLYMO (2.9) NPs. The solution remained clear during the measurement, indicating colloidal stability of the NP-surfactant mixture. The pure $CO_2$-API brine interfacial tension was $33.11 \pm 0.62$ mN/m. (b) Surface tension at RT and ambient pressure of Neodol 25-9 in DI water (red squares) and API brine without (blue diamonds) or with (green triangles) 1% PEG(6-9EO) (1.7) NPs. The pure DI-air surface tension was 72.52 mN/m and the pure API-air surface tension was 75.16. The API-air surface tension with only 1% PEG(6-9EO) (1.7) NPs was 54.43 mN/m and 1% GLYMO (2.9) NPs was 69.54 mN/m. NP dispersions were at pH~8.5.

Table S2. Tabulated IFT data from Figure S3.

| NP | Surfactant | Conditions | IFT value± Std. Dev. (mN/m) |
|---|---|---|---|
| none | none | C-W, RT, 2800 psia, API brine | 33.11±0.62 |
| none | 0.001% AOS12 | C-W, RT, 2800 psia, API brine | 28.89±1.09 |
| none | 0.003% AOS12 | C-W, RT, 2800 psia, API brine | 12.11±0.95 |
| none | 0.01% AOS12 | C-W, RT, 2800 psia, API brine | 5.80±0.20 |
| none | 0.03% AOS12 | C-W, RT, 2800 psia, API brine | 4.32±0.07 |
| none | 0.1% AOS12 | C-W, RT, 2800 psia, API brine | 3.90±0.17 |
| none | 0.3% AOS12 | C-W, RT, 2800 psia, API brine | 4.32±0.43 |
| 1% GLYMO (2.9) | 0.001% AOS12 | C-W, RT, 2800 psia, API brine | 25.49±3.57 |
| 1% GLYMO (2.9) | 0.01% AOS12 | C-W, RT, 2800 psia, API brine | 6.93±0.21 |
| 1% GLYMO (2.9) | 0.1% AOS12 | C-W, RT, 2800 psia, API brine | 3.43±0.59 |
| none | none | Air-water, RT, atmospheric pressure, DI water | 72.52 |
| none | 0.0001% Neodol 25-9 | Air-water, RT, atmospheric pressure, DI water | 70.66 |
| none | 0.001% Neodol 25-9 | Air-water, RT, atmospheric pressure, DI water | 56.44 |
| none | 0.002% Neodol 25-9 | Air-water, RT, atmospheric pressure, DI water | 44.17 |
| none | 0.005% Neodol 25-9 | Air-water, RT, atmospheric pressure, DI water | 34.95 |
| none | 0.01% Neodol 25-9 | Air-water, RT, atmospheric pressure, DI water | 32.18 |
| none | 0.1% Neodol 25-9 | Air-water, RT, atmospheric pressure, DI water | 30.63 |
| none | 1% Neodol 25-9 | Air-water, RT, atmospheric pressure, DI water | 29.79 |
| none | none | Air-water, RT, atmospheric pressure, API brine | 75.16±1.15 |
| none | 0.0001% Neodol 25-9 | Air-water, RT, atmospheric pressure, API brine | 73.19±0.99 |
| none | 0.001% Neodol 25-9 | Air-water, RT, atmospheric pressure, API brine | 54.48±0.35 |
| none | 0.003% Neodol 25-9 | Air-water, RT, atmospheric pressure, API brine | 34.62±1.48 |
| none | 0.01% Neodol 25-9 | Air-water, RT, atmospheric pressure, API brine | 31.69±0.88 |
| none | 0.035% Neodol 25-9 | Air-water, RT, atmospheric pressure, API brine | 31.28±0.30 |
| none | 0.25% Neodol 25-9 | Air-water, RT, atmospheric pressure, API brine | 29.30±0.46 |
| none | 0.5% Neodol 25-9 | Air-water, RT, atmospheric pressure, API brine | 28.96±0.11 |
| none | none | Air-water, RT, atmospheric pressure, API brine | 54.43 |
| 1% PEG(6-9EO) (1.7) | 0.0001% Neodol 25-9 | Air-water, RT, atmospheric pressure, API brine | 54.45 |
| 1% PEG(6-9EO) (1.7) | 0.001% Neodol 25-9 | Air-water, RT, atmospheric pressure, API brine | 53.96 |
| 1% PEG(6-9EO) (1.7) | 0.003% Neodol 25-9 | Air-water, RT, atmospheric pressure, API brine | 35.89 |
| 1% PEG(6-9EO) (1.7) | 0.01% Neodol 25-9 | Air-water, RT, atmospheric pressure, API brine | 30.35 |
| 1% PEG(6-9EO) (1.7) | 0.1% Neodol 25-9 | Air-water, RT, atmospheric pressure, API brine | 28.55 |

The presence of NPs in surfactant formulations may provide additional performance and functionality, including improved emulsion/foam stability,[1] "tracer" capabilities for tracking flow,[54] magnetic response for imaging,[43-44] catalytic behavior for environmental remediation,[55] wettability alteration of mineral surfaces for EOR,[56] and/or blocking of pores for control of fluid placement in porous media. In contrast to the many emerging capabilities of NPs, the roles of surfactants are well established, and as such, the following discussion is organized by the type of surfactant combined with the NPs to illustrate possible benefits NPs can add to established surfactant types (anionic, nonionic, and finally cationic). Novel aspects of these new NP + surfactant formulations will be discussed below to provide a basis for future studies.

Table S3. Viscosity summary of NP + surfactant foams formed in 22D beadpack at 3mL/min at 2800 psia. The numbers in parentheses in the last two columns are the values obtained for the surfactant with the added nanoparticles.

| Surfactant | NP | Salinity | T, °C | Quality | $\mu_{app, beadpack}$ (with NP), cP | $\mu_{app, capillary}$ (with NP), cP |
|---|---|---|---|---|---|---|
| 0.05% AOS12 | 1% GLYMO (1.8) | API brine | RT | 75% | (14.8) | (10.3) |
| 0.05% AOS12 | 1% GLYMO (2.9) | API brine | RT | 95% | 2.0 (1.9) | 0.1 (0.4) |
| 0.05% AOS12 | 1% PEG(6-9EO) (2.1) | API brine | RT | 76% | (5.1) | (2.5) |
| 0.05% AOS12 | 1% PEG(6-9EO) (2.1) | API brine | RT | 95% | 2.0 (2.6) | 0.1 (0.8) |
| 0.3% AOS12 | 1% GLYMO (2.9) | API brine | RT | 75% | 11.5 (10.3) | 11.8 (11.6) |
| 0.3% AOS12 | 1% GLYMO (2.9) | API brine | RT | 95% | 11.1 (17.4) | 10.8 (15.3) |
| 0.03% AOS14-16 | 0.25% SB | 4% NaCl | RT | 75% | 11.9 (9.2) | 4.4 (3.8) |
| 0.03% AOS14-16 | 1% SB | 4% NaCl | RT | 75% | 11.9 (12.1) | 4.4 (3.0) |
| 4% SLES | 1% GLYMO (2.9) | API brine | RT | 95% | 27.3 (20.1) | >52.8 (52.0) |
| 4% SLES | 1% SB | API brine | RT | 95% | 27.3 (25.2) | >52.8 (42.9) |
| 0.01% N25-9 | 1% NexSil6+PEG(6-9EO) (1.7) | API brine | 43 | 75% | (1.4) | (2.0) |
| 0.05% N25-9 | 1% PEG(6-9EO) (2.1) | API brine | RT | 75% | 18.1 (17.7) | 22.6 (21.4) |
| 0.05% N25-9 | 0.3% PEG(6-9EO) (2.1) | API brine | RT | 75% | 18.1 (16.2) | 22.6 (21.4) |
| 0.05% N25-9 | 0.3% PEG(6-9EO) (1.7) | API brine | 43 | 75% | (1.9) | (1.3) |
| 0.5% L24-22 | 1% GLYMO (2.9) | API brine | RT | 95% | 22.2 (19.8) | 19.1 (18.5) |
| 0.01% N-300 | 1% GLYMO (2.9) | API brine | RT | 75% | 13.8 (10.7) | 2.8 (2.1) |
| 0.02% N-300 | 1% GLYMO (2.9) | API brine | RT | 75% | 13.7 (14.3) | 3.9 (2.6) |
| 0.05% N-300 | 1% GLYMO (2.9) | API brine | RT | 75% | 21.5 (18.2) | 19.0 (19.1) |
| 0.5% N-300 | 1% GLYMO (2.9) | API brine | RT | 95% | 14.7 (15.8) | 20.9 (27.4) |
| 0.5% N-300 | 1% PEG(8-12EO) | API brine | RT | 75% | 16.0 (14.6) | 25.3 (20.9) |
| 0.75% N-300 | 1% GLYMO (2.9) | API brine | RT | 95% | 17.9 (25.4) | 36.9 (43.8) |
| 0.5% N-300 | 1% GLYMO (2.6) | API brine | RT | 95% | (28.65) | (24.4) |
| 0.5% C12TAB | 1% GLYMO (2.9) | API brine | RT | 95% | 18.2 (18.6) | 23.3 (13.8) |
| 0.5% C12TAB | 1% PEG(6-9EO) (2.1) | API brine | RT | 95% | 18.2 (7.3) | 23.3 (3.9) |
| 0.5% C16TAB | 1% GLYMO (2.9) | ¼ API | RT | 95% | 12.2 (10.3) | 7.1 (19.2) |
| 0.5% C16TAB | 1% GLYMO (2.9) | API brine | RT | 95% | 13.1 (14.4) | 10.9 (9.4) |
| 0.5% C16TAB | 1% PEG(6-9EO) (2.1) | API brine | RT | 95% | 13.1 (12.4) | 10.9 (7.9) |

**Zeta potential measurement**

At low pH silica nanoparticles are more protonated, whereas at higher pH the zeta potential of the silica nanoparticles reached a plateau at about pH ~ 9. The PEG (6-9)EO (2.1) nanoparticles showed the lowest surface charge in this measurement.

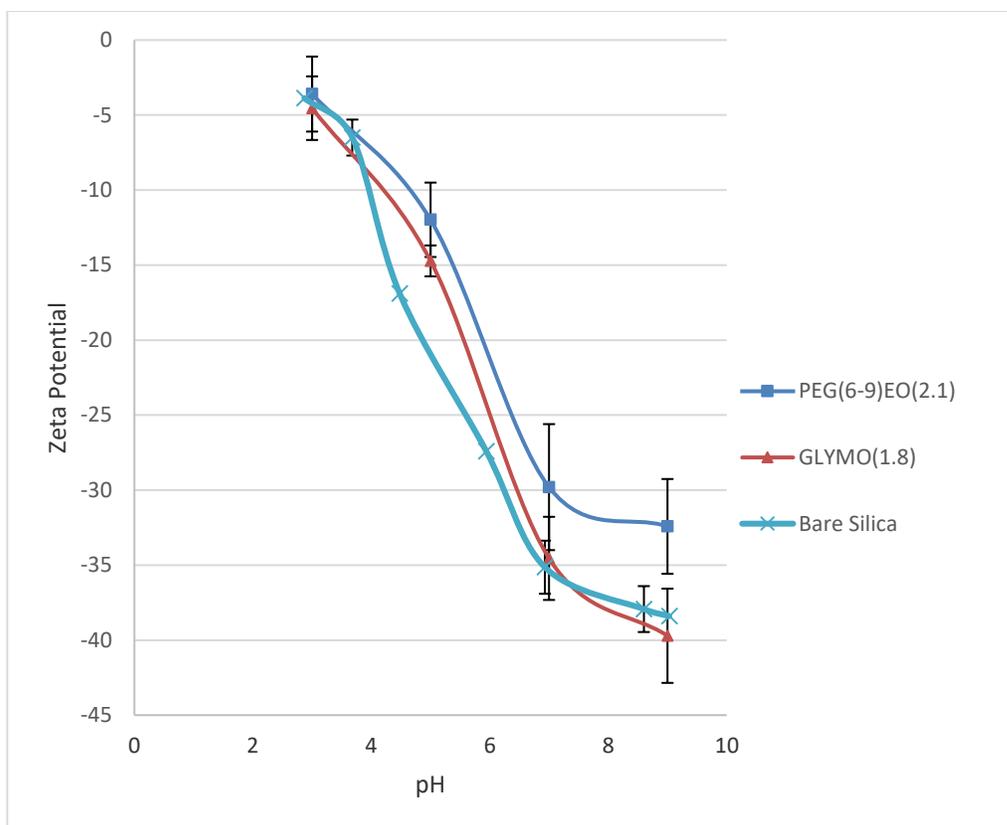

Figure S4: Zeta Potential vs. pH of silica nanoparticles with 10 mM KCl added, collected with Brookhaven ZetaPALS (open symbols).

## C/W foam stability determination

The foam bubble size and stability were measured in situ with a high pressure microscope cell installed at the exit of the beadpack.[8, 41] A Nissan Eclipse ME600 microscope equipped with a Photometrics CoolSNAP CF CCD camera was used to collect images of foam in the microscope cell (Figure S5 and Figure 2 insets). For long term measurement of foam stability over several hundred minutes (Figure S6 and Table S5), foams were sealed in the cell by closing a three-way valve at both the inlet and outlet. At least 100 bubbles were analyzed at each condition using Image J software.

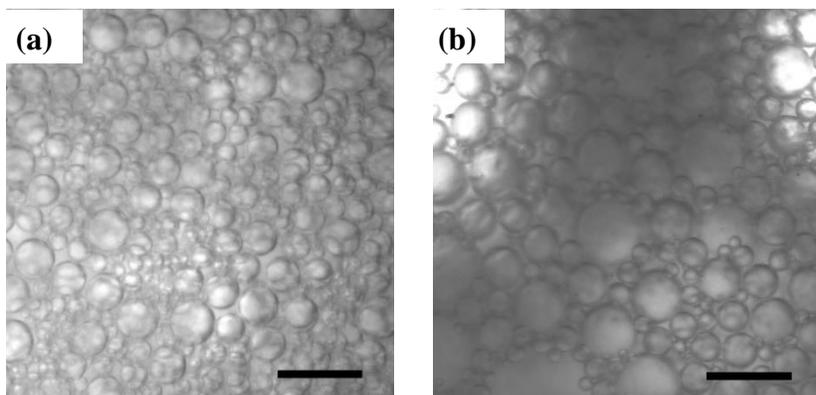

Figure S5. Optical micrographs of foams stabilized by PEG(6-9EO) (2.1)-coated silica NPs in SSW at RT, 2800 psia, and a quality of 75% at (a) t=0 (b) t=120 min.

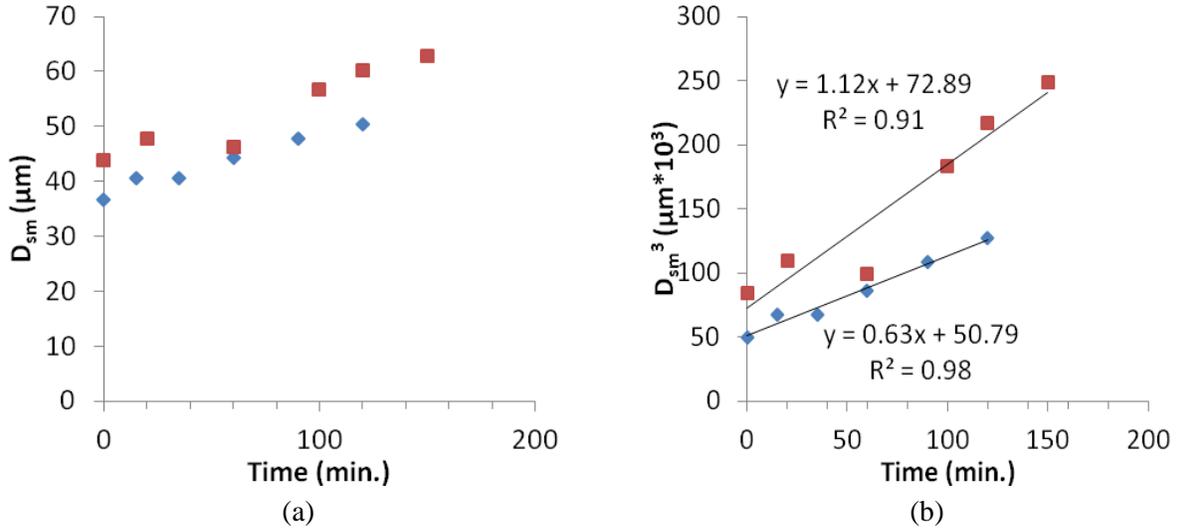

(a) (b)

Figure S6. (a) Sauter mean diameter (Dsm) vs. time of 75% quality C/W foam stabilized by 1% PEG(6-9EO) (2.1) NPs in SSW (diamonds) and API brine (squares) at RT and 2800. (b) $D_{sm}^3$ vs. time and linear regression fits of the data from (a). The initial $D_{sm}$ was 37µm in SSW and 44µm in API brine.

Figures S5 illustrates the evolution of $CO_2$ bubbles stabilized by PEG(6-9EO)-coated silica NPs in SSW or API brine observed in the high pressure microscope cell. The foam bubbles become notably larger after 120 min. To quantify this change in size, the Sauter mean diameter ($D_{sm}$) was determined and is plotted vs. time in Figure S6a. In SSW, the $D_{sm}$ increased from 37 µm to 50 µm in 120 minutes. In API brine, the $D_{sm}$ increased from 44 µm to 60 µm in the same time.

According to Lifshitz, Slyozov, and Wagner (LSW) theory the Ostwald ripening rate, $\Omega_3$, is given by

$$\Omega_3 = \frac{dD_{sm}^3}{dt} = \frac{64\gamma D_{diff} SV_m}{9RT} F \qquad (1)$$

where $D_{diff}$ is the molecular diffusion coefficient, S is the bulk solubility of the dispersed phase, $V_m$ is the molar volume of the dispersed phase, and F is a correction factor to account for the diffusion length at a given quality. To experimentally determine $\Omega_3$, the data given in Figure S6a were linearized by plotting $D_{sm}^3$ vs. time in Figure S6b and best fit lines are given. Thus, the slope of the line is the estimated $\Omega_3$ value whereby lower values indicate more stable foams. As indicated in Figure S6b, $\Omega_3$ was slightly lower in SSW than in API brine, possibly due to the higher uniformity of bubble size in SSW (Figure 8.5). The $\Omega_3$ values of 0.6~1.1 µm$^3$*10$^3$/min.

for NP-stabilized foams are between those collected at 3000 psia and RT for foams of 95% quality with 0.08% lauramidopropyl betaine (LAPB) surfactant + 1% proprietary NPs with ($\Omega_3$ = 0.5 µm$^3$*10$^3$/min.) and without ($\Omega_3$ = 3.3 µm$^3$*10$^3$/min.) 0.88% partially hydrolyzed polyacrylamide (HPAM) generated at the same flow conditions.[8] It is also notable that the NP-only foams presented in the present work were more stable than foams generated without NPs and only 0.88% HPAM + 0.08% LAPB ($\Omega_3$ = 16.6 µm$^3$*10$^3$/min.).[8]

Following the methodology used to produce Figure S6b, Table S4 summarizes the initial Dsm values and experimental Ω3 for foams formed in a 22D beadpack at 3mL/min at 2800 psia with mixtures of NPs and surfactants. In the remainder of this supporting information, the viscosity (Table S3) and stability data from microscopy (Table S4) will be discussed for each class of surfactants.

Table S4. Stability summary of NP + surfactant foams formed in 22D beadpack at 3mL/min at 2800 psia. See Table S3 for further explanation.

| Surfactant | NP | Salinity | T, °C | Quality | $D_{sm}$ at t=0 (with NPs), μm | Experimental $\Omega_3$ (with NPs), $\mu m^3 * 10^3$/min. |
|---|---|---|---|---|---|---|
| 0.3% AOS12 | 1% GLYMO (2.9) | API brine | RT | 95% | 37 (29) | 5.3 (12.0) |
| 0.3% AOS12 | 1% GLYMO (2.9) | API brine | RT | 75% | 41 (43) | 1.0 (1.9) |
| 4% SLES | 1% GLYMO (2.9) | API brine | RT | 95% | 21 (27) | 0.7 (1.0) |
| 0.05% N25-9 | 1% PEG(6-9EO) | API brine | RT | 75% | 70 (57) | 4.1 (7.0) |
| 0.5% L24-22 | 1% GLYMO (2.9) | API brine | RT | 95% | 40 (37) | 4.8 (3.4) |
| 0.5% N-300 | 1% GLYMO (2.9) | API brine | RT | 95% | 33 (25) | 2.6 (0.2) |
| 0.5% N-300 | 1% GLYMO (2.6) | API brine | RT | 95% | 33 (26.4) | 2.6 (1.8) |
| 0.5% N-300 | 1% GLYMO (2.9) | API brine | 50 | 95% | 36 (41) | 15 (8.7) |
| 0.05% N-300 | 1% GLYMO (2.9) | API brine | RT | 75% | 39 (38) | 0.4 (0.3) |
| 0.5% N-300 | 1% GLYMO (2.9) | API brine | RT | 75% | 40 (37) | 0.8 (0.8) |
| 0.5% C16TAB | 1% GLYMO (2.9) | ¼ API | RT | 95% | 41 (27) | 8.2 (0.9) |
| 0.5% C16TAB | 1% GLYMO (2.9) | API brine | RT | 95% | 80 (37) | 7.9 (10.0) |
| 0.5% C16TAB | 1% PEG(6-9EO) (2.1) | API brine | RT | 95% | 80 (38) | 7.9 (0.6) |

**NP+surfactant formulations for C/W foams**

**Anionic surfactant formulations.** Tables S3 and S4 give three examples of anionic surfactants which each generated viscous C/W foam without added NPs: AOS12, AOS14-16, and SLES. Anionic surfactants are of interest for many applications, including EOR due to their low adsorption on anionic sandstone and have been used in commercial C/W foam pilot studies.[57-58] Anionic surfactants are known to synergistically interact with zwitterionic surfactants[5] and free PEG molecules,[59] but are rarely combined with functionalized NPs with various ligand types grafted to their surfaces.

At a low concentration of 0.05% AOS12, viscous foams were not formed at a high quality of 95%, but were formed at a lower quality of 75% (Table S3). Foam formation may have been improved at the lower quality due to the lower capillary pressure[60] and due to the higher overall concentration of surfactant in the total mixture of aqueous phase plus $CO_2$ phase. Interestingly, when 1% GLYMO (1.8)- or PEG(6-9EO) (2.1)-coated NPs were added, the viscosity of the resulting foam was somewhat lower than for the NPs only. When slightly-interfacially active 1% GLYMO (2.9)-coated NPs were added, no such competition was observed and the foam behaved as the surfactant only. The lack of competition with slightly-interfacially active GLYMO-coated NPs suggests there is negligible interaction between the NPs and the surfactant. At a higher AOS12 concentration of 0.3%, foams with a high viscosity of 10.8~11.8 cP were formed in both the beadpack and the capillary at qualities of 75% and 95% (Table S3). However, the stability of foams formed with 0.3% AOS12 was lower at the higher quality of 95% that at 75% (Table S4), again likely influenced by the capillary pressure and overall surfactant concentration in the mixture. When 1% of non-interfacially active GLYMO (2.9)-coated NPs were added, no significant improvement was observed in either foam viscosity (Table S3) or stability (Table S4). When the longer-tailed AOS14-16 was investigated, we noted that the surfactant was insoluble in API brine and thus a lower salinity of 4% NaCl was chosen. Table S3 indicates that 0.03% AOS14-16 in 4% NaCl made foam with viscosity of 11.9cP in the beadpack, very similar to 0.3% AOS12 in API brine. When 0.25% or 1% non-interfacially active SB-coated NPs were added, no significant change in foam viscosity was observed compared to the surfactant-only case (Table S3).

When 4% of the viscoelastic SLES surfactant was used in API brine, high viscosity foams (27.3 cP in the beadpack and >52.8 cP in the capillary tube) were formed at the high quality of 95%, in agreement with previously reported results.[61] Note that the lower limit of ">52.8 cP" is given in the capillary tube because the 100 psi differential pressure transducer was reading its full scale during the experiment, indicating an unmeasurable pressure drop of over 100 psi in the capillary tube. Entanglement of the wormlike micelles of SLES can significantly increase the viscosity of the bulk aqueous phase and thus improve the foam viscosity.[61] The extremely high viscosities measured in the capillary tube are likely due to the shear thinning nature of the viscoelastic wormlike micelles in the foam[61] and lower shear of the capillary compared to the beadpack. Similarly to the AOS12 and AOS14-16 experiments, when slightly-interfacially active NPs (1% GLYMO (2.9)- or SB- coated NPs) were added to the surfactant formulation, no significant change in viscosity (Table S3) or stability (Table S4) was observed. The lack of significant influence of the surface-modified NPs on the performance of SLES suggested that the NPs did not significantly affect the interfacial or micellar behavior of the surfactant.

**Nonionic surfactant formulations.** Tables S3 and S4 give three examples of nonionic surfactants which each could generate viscous C/W foam without added NPs: N25-9, L24-22, and N-300. Nonionic surfactants tend to be lower in cost than ionic surfactants, which makes them desirable for many applications. For subsurface applications, however, nonionic surfactants tend to adsorb more strongly on sandstone surfaces than anionic surfactants, which may act to balance the advantage of lower cost somewhat, but they have still been used in commercial C/W foam EOR pilot studies.[57, 62] The interaction between nonionic surfactants and sandstone (silica) is primarily due to H-bonding between ether oxygen groups on the surfactants (H-bond acceptors) and the SiOH groups on the silica (H-bond donators). This same interaction is what often causes both synergy and destabilization of bare silica NPs in the presence of free nonionic surfactants.[25, 49] As demonstrated in Table 5, the coatings used in this study prevented aggregation of silica NPs in the presence of free nonionic surfactants, allowing new formulations of NPs and surfactants for C/W foam formation. Further, the lack of charge of nonionic surfactants is often favorable for formulations because they lack appreciable electrostatic interaction between other species to maximize compatibility. To balance the requirements of interfacial activity with solubility in high salinity brine (e.g. API brine where solvation of nonionic surfactants by H-bonding with water is weakened), nonionic surfactants with long EO chains have been demonstrated recently[63] and are investigated in the present study.

As shown in Tables S3 and S4, N25-9 is remarkable for generating foams which were highly viscous (18.1 cP in the beadpack and 21.4 cP in the capillary) and moderately stable ($\Omega_3$ = 4.1) with only 0.05% surfactant alone at a quality of 75%. In contrast to a decrease in foam viscosity observed when interfacially active NPs were added to 0.05% AOS12 (discussed above), the addition of 1% PEG(6-9EO)-coated NPs to N25-9 did not significantly change the viscosity (Table S3) or stability (Table S4) of the foam compared to the surfactant-only case. Foam stabilized by 0.5% of the structurally similar L24-22 surfactant was likewise unaffected by the addition of GLYMO (2.9)-coated NPs (Tables S3 and S4). The foam stabilized by 0.5% L24-22 was notable given the high viscosity of ca. 20 cP in both the beadpack and capillary tube at a high quality of 95% (Table S3), potentially making it of interest for applications such as hydraulic fracturing fluids.[8, 61]

N-300, a nonylphenolic nonionic surfactant of potential interest for C/W foam pilot tests,[64] generated some of the highest reported bulk foam (i.e. in the capillary tube) viscosities (36.9 cP) for a surfactant-only system which was not viscoelastic.[8, 41, 61] When 1% GLYMO (2.9)-coated NPs were added with 0.01-0.75% N-300, no significant change was observed in terms of foam viscosity at the conditions tested (Table S3). Interestingly, the stability of the C/W foam was improved at RT ($\Omega_3$ decreased by factor of 13, Table S4) but minimal change was evident at 50°C (Table S4).

**Cationic surfactant formulations.** Cationic surfactants are of interest for many applications, and are of interest for EOR in positively charged carbonate reservoirs.[65-68] Tables S3 and S4 give two examples of C/W foams formed with quaternary amine cationic surfactants \ combined with surface-modified silica NPs. The positive charge on the quaternary amines gives high solubility in brine over a wide pH range. The positive charge also causes very strong interaction with bare sandstone (silica) surfaces, promoting adsorption. The same electrostatic attraction is responsible for both synergy and destabilization of bare silica NPs in the presence of free cationic surfactants,[26] similar to the case of nonionic surfactants discussed above. Indeed, many examples of quaternary amines used with bare silica at low or zero salinity are given in the literature.[26, 69] Importantly, the coatings presented in this study provided stabilization of silica NPs in the presence of free quaternary amine surfactants (Table 5), allowing new formulations to be tested for C/W foam formation.

The cationic surfactants C12TAB and C16TAB, differing by 4 carbon atoms in the hydrophobic tail, were combined with GLYMO (2.9)- and PEG(6-9EO) (2.1)-coated NPs and the resulting C/W foam properties are reported in Tables S3 and S4. When combined with 1% GLYMO (2.9)-coated NPs, no significant change in foam viscosity was observed compared to the surfactant only cases (Table S3). The stability of foam stabilized by 0.5% C16TAB was likewise unaffected in API brine with the addition of 1% GLYMO (2.9)-coated NPs (Table S4), but $\Omega_3$ decreased by a factor of 9 in ¼ API brine, indicating improved foam stability where electrostatic attraction between the NPs and surfactant would be less screened. When 0.5% C12TAB was combined with 1% PEG(6-9EO) (2.1)-coated NPs in API brine, a significant decrease in foam viscosity was observed, consistent with the anionic and nonionic surfactant data reported in Table S3 and discussed above. At the same conditions, an increase in stability by a factor of 13 to $\Omega_3 = 0.6$ was observed with the addition of 1% PEG(6-9EO) (2.1)-coated NPs (Table S4), giving a foam stability very similar to the NP only case of $\Omega_3 = 1.1$ (Figure S6). The authors caution that the stability points showing significant improvement in foam stability were each only tested once, indicating that further testing is warranted.